\documentclass{article}

\newcommand{\mI}{\boldsymbol{I}}
\newcommand{\mJ}{\boldsymbol{J}}
\newcommand{\mK}{\boldsymbol{K}}
\newcommand{\mr}{\boldsymbol{r}}
\newcommand{\ms}{\boldsymbol{s}}
\newcommand{\mt}{\boldsymbol{t}}
\newcommand{\mpp}{\boldsymbol{p}}
\newcommand{\mq}{\boldsymbol{q}}
\newcommand{\ma}{\boldsymbol{a}}
\newcommand{\mb}{\boldsymbol{b}}
\newcommand{\malpha}{\boldsymbol{\alpha}}
\newcommand{\mbeta}{\boldsymbol{\beta}}
\newcommand{\mgamma}{\boldsymbol{\gamma}}

\newcommand{\p}[1]{(\ref{#1})}

\newcommand{\cF}{{\cal F}}

\newcommand{\cD}{{\cal D}}

\newcommand{\cX}{{\cal X}}

\newcommand{\be}{\begin{equation}}
\newcommand{\ee}{\end{equation}}
\newcommand{\bea}{\begin{eqnarray}}
\newcommand{\eea}{\end{eqnarray}}

\newcommand{\ba}{\begin{array}} \newcommand{\ea}{\end{array}}

\def\im{{\rm i}}

\newcommand{\nn}{\nonumber}

\usepackage{amscd,amsmath,amssymb}

\begin{document}
\thispagestyle{empty}
\vspace{2cm}
\begin{flushright}
\end{flushright}\vspace{2cm}
\begin{center}
{\Large\bf Improved non-Abelian tensor multiplet action}
\end{center}
\vspace{1cm}

\begin{center}
{\large\bf  N.~Kozyrev${}^a$}
\end{center}

\begin{center}
${}^a$ {\it
Bogoliubov  Laboratory of Theoretical Physics, JINR,
141980 Dubna, Russia} \vspace{0.2cm}
\end{center}
\vspace{2cm}

\begin{abstract}\noindent
Construction of the superfield action of the $N=(1,0)$, $d=6$ non-Abelian tensor multiplet based on the non-Abelian tensor hierarchies is considered. It is shown that while straightforward non-Abelian generalization of the Pasti-Sorokin-Tonin action is not a workable solution, a suitable truncation of the PST action can be still be modified to include non-Abelian tensor field. In the modified action, the self-dual equation of motion of the tensor field is induced by a composite Lagrange multiplier, which is not a component of a standard dynamical tensor multiplet. As a result, the constraint that enforces the gauge group to be non-compact, as in the usual tensor hierarchies, can be avoided.
\end{abstract}

\setcounter{page}{1}
\setcounter{equation}{0}


\section{Introduction}
The non-Abelian tensor gauge theories have been under consideration for a long time. One of modern motivations to study such theories comes from $M-$theory, as the low effective description of one of its extended objects, $M5$-brane, is given by the Born-Infeld type tensor multiplet action \cite{M5brane}, and multiple $M5$-branes are expected to be described a non-Abelian version of such a theory \cite{BLM}. Another point of interest is the fact that maximum possible superconformal theory is a six-dimensional theory with $N=(2,0)$ supersymmetry \cite{witten}, and its degrees of freedom should combine into the tensor multiplet. Another factors that make this theory mysterious are arguments by Witten based on dimensional reduction that an action with conformal symmetry should not exist \cite{witten2}, and results by Bekaert, Henneaux, Knaepen and Sevrin \cite{henneaux1,bekaert1}   that usual tensor field gauge symmetry $\delta B_{MN} = \partial_M a_N - \partial_N a_M$ can not be deformed in the non-Abelian way, at least if theory is local and no other fields are present. \footnote{More issues with Lagrangian description of multiple $M5$-branes can be found in \cite{lagr20} and references therein.}

Many approaches were proposed so far to construct an action that would reflect at least some of the properties of the non-Abelian tensor theory. Among them are introduction of non-localities \cite{nonloc}, attempts to extend a five-dimensional Yang-Mills theory \cite{gustavsson}, construction of tensor self-dual ``field strength'' in terms of Yang-Mills strength tensor and a covariantly constant vector \cite{lambert1} \footnote{This approach is related to 3-algebras which appear three-dimensional BLG theory \cite{BL1},\cite{gustavsson2}.}, and more. The most conventional approach seems to be the one based on so-called tensor hierarchies, which were introduced in \cite{tenshier1} and combined with $N=(1,0)$ supersymmetry using standard fields in \cite{tenshier2}, although some problems were noticed. In particular, it was found that in many cases theory can be formulated only in terms of equations of motion, and the action exists only when the metric in the internal sector is indefinite (and the gauge group is non-compact). Also, these systems in general possess a cubic potential for scalar fields, which is not positive-definite and thus should not be compatible with supersymmetry.

The tensor hierarchy actions were elaborated further in \cite{tenshier3}, where explicit solutions for hierarchy structure constants in terms of usual Lie algebra representations and tensors were found. Superfield version of the tensor hierarchy was constructed in \cite{bandos}, and the harmonic superspace action in \cite{buchbinder}. Characteristic feature of these constructions is the presence of two different dynamical tensor multiplets, one of which contains a two-form potential and the other self-dual ``field strength'', which induces the self-dual equation of motion of the latter. Scalar kinetic term contains fields of both multiplets linearly, and, therefore, possesses indefinite metric.

A possibility to avoid presence of two multiplets and resulting indefiniteness of the metric is the use of Past-Sorokin-Tonin mechanism of generating self-dual equation of motion for the tensor field, which was done in \cite{BSS}. This construction, however, uses the tensor hierarchy with the same constraints on the coupling constants, thus being unable to solve the problem of indefiniteness of the metric by itself. Also, the 3-form is still present and appears to be required to prove invariance of the action with respect to PST gauge symmetries.

In this article, we discuss the possibility of construction of the non-Abelian tensor multiplet action with positive-definite metric in the scalar sector. Firstly, we consider the bosonic tensor hierarchy \cite{tenshier2,tenshier3} to show that it can be treated, without substantial loss of generality, as a system of tensor fields without standard shift symmetries coupled to the standard Yang-Mills field. Secondly, we discuss simple non-Abelian deformation of the PST action to show that it does not possess the required symmetries and does not generate a useful equation of motion without an additional 3-form present. Thirdly, we consider an alternative approach based on the truncation of the PST action that is equivalent to the standard Lagrange multiplier method in the purely bosonic case but allows for different supersymmetrization. Finally, we present a manifestly supersymmetric harmonic superspace action and show that it possesses the required bosonic limit without putting unappropriate constraints on the gauge group and briefly discuss its properties.

Throughout this article, we use vector and spinor notations interchangeably, with the details given in the Appendix A. Standard and capital letters, Greek or Latin, are reserved for space-time and harmonic indices, while bold ones are used for internal symmetries.


\section{Preliminary considerations}
\subsection{The tensor hierarchy}
Among a number of approaches to the non-Abelian tensor theories, the tensor hierarchies approach seems to be most conventional. It was introduced in \cite{tenshier1} and applied to component theories with $N=(1,0)$, $d=6$ supersymmetry in \cite{tenshier2}. Bosonic gauge transformations minimally involve fields $A_M^{\mr}$, $B_{[MN]}^{\mI}$, $C_{[MNK]\mr}$:
\bea\label{tenshier}
\delta A^{\mr}_M &=& \cD_M \Lambda^{\mr} - h^{\mr}_{\mI} \Lambda^{\mI}_M, \nn \\
\delta B^{\mI}_{[MN]} &=& 2 \cD_{[M} \Lambda^{\mI}_{N]} -2 d^{\mI}_{\mr\ms} \Lambda^{\mr} F^{\ms}_{MN} +2d^{\mI}_{\mr\ms} A^{\mr}_{[M}\delta A^{\ms}_{N]} - g^{\mI\mr} \Lambda_{MN\mr}, \nn \\
\delta C_{[MNK]\mr} &=& 3 \cD_{[M}\Lambda_{NK]\mr}  + 3 b_{\mI\mr\ms} F^{\ms}_{[MN} \Lambda^{\mI}_{K]} + b_{\mI\mr\ms} F^{\mI}_{MNK}\Lambda^{\ms} + \\
&&+3 b_{\mI\mr\ms} B^{\mI}_{[MN} \delta A^{\ms}_{K]} +2 b_{\mI\mr\ms} d^{\mI}_{\mpp\mq} A^{\ms}_{[M}A^{\mpp}_{N}\, \delta A^{\mq}_{K]}+\ldots\nn
\eea
Here $F^{\mr}_{MN}$ and $F^{\mI}_{MNK}$ are covariant field strengths
\bea\label{hierfieldstr}
F^{\mr}_{MN} &=& 2 \partial_{[M} A^{\mr}_{N]} - f^{\mr}_{\ms\mt} A^{\ms}_M \, A^{\mt}_{N} + h^{\mr}_{\mI} B^{\mI}_{MN}, \nn \\
F^{\mI}_{MNK} &=& 3 \cD_{[M} B^{\mI}_{NK]} +6 d^{\mI}_{\mr\ms} A^{\mr}_{[M}\,  \partial_N A^{\ms}_{K] } -2 f^{\ms}_{\mpp\mq} d^{\mI}_{\mr\ms} A^{\mr}_{[M} A^{\mpp}_N A^{\mq}_{K] } + g^{\mI\mr} C_{MNK\mr}
\eea
and $f_{\mr\ms}^{\mt}$, $d^{\mI}_{\mr\ms}$, $b_{\mI\mr\ms}$, $h^{\mr}_{\mI}$, $g^{\mI\mr}$ are dimensionless coupling constants. They satisfy a system of constraints which imply \cite{tenshier2} that composite quantities
\be\label{tensgens}
\big( T_{\mr}   \big)^{\mt}{}_{\ms} = - f_{\mr\ms}^{\mt} + d^{\mI}_{\mr\ms} h^{\mt}_{\mI}, \;\; \big( T_{\mr}  \big)^{\mJ}{}_{\mI} = 2 h^{\ms}_{\mI} d^{\mJ}_{\mr\ms} - g^{\mJ\ms}b_{\mI\ms\mr}
\ee
form a Lie algebra
\bea\label{lie}
&\big( T_{\mr}  \big)^{\mt}{}_{\mq} \big( T_{\ms}  \big)^{\mpp}{}_{\mt} -\big( T_{\ms}  \big)^{\mt}{}_{\mq} \big( T_{\mr}  \big)^{\mpp}{}_{\mt} = f_{\mr\ms}^{\mt} \big( T_{\mt}  \big)^{\mpp}{}_{\mq},& \nn \\
&\big( T_{\mr}  \big)^{\mJ}{}_{\mI} \big( T_{\ms}  \big)^{\mK}{}_{\mJ} - \big( T_{\ms}  \big)^{\mJ}{}_{\mI} \big( T_{\mr}  \big)^{\mK}{}_{\mJ} = f_{\mr\ms}^{\mt} \big( T_{\mt}  \big)^{\mK}{}_{\mI}&
\eea
with coupling constants being its invariant tensors. $T_{\mr}$ are used to define covariant derivatives
\be\label{tenscovder}
\cD_M = \partial_M - A_M^{\mr} T_{\mr} \cdot, \;\; T_{\mr}\cdot A^{\ms} = - \big( T_{\mr}   \big)^{\ms}{}_{\mt} A^{\mt}, \;\;  T_{\mr}\cdot A^{\mI} =- \big( T_{\mr}  \big)^{\mI}{}_{\mJ}A^{\mJ}.
\ee
In \cite{tenshier2}, two fermionic fields, $\chi^{i\mI}_{\alpha}$ and $\lambda^{i\mr\alpha}$, were added and $N=(1,0)$ supersymmetry transformations  found. Their algebra closes on rather complex equations of motion; it was shown in \cite{tenshier2} that they can be integrated to the Lagrangian only if additional relations hold
\be\label{lagrcond}
h^{\mr}_{\mI} = \eta_{\mI\mJ}g^{\mJ\mr}, \;\; b_{\mI\mr\ms} = 2 \eta_{\mI\mJ}d^{\mJ}_{\mr\ms}, \;\; \eta_{\mI\mJ}d^{\mI}_{\mr(\ms}d^{\mJ}_{\mpp\mq)}=0.
\ee
Here, $\eta_{\mI\mJ}$ is a non-degenerate metric. The first of these relations indicates a problem, however, as to satisfy also one of consistency relations  $h^{\mr}_{\mI}g^{\ms\mI}=0$, $\eta^{\mI\mJ}$ must be indefinite. Also, for the systems with Lagrangian neither $h^{\mr}_{\mI}$ or $g^{\ms\mI}$ can be zero, suggesting that the presence of the 3-form is necessary.

Important observation about the tensor hierarchy comes from two consistency equations
\bea\label{rels12}
f_{\mr\ms}^{\mt}h^{\mr}_{\mI} - d^{\mJ}_{\mr\ms}h^{\mt}_{\mJ} h^{\mr}_{\mI} =0 \; \Rightarrow \; h^{\mr}_{\mI} \big(T_{\mr}\big)^{\mt}{}_{\ms}=0, \nn \\
g^{\mJ\ms} h^{\mr}_{\mK} b_{\mI\ms\mr} - 2 h^{\ms}_{\mI} h^{\mr}_{\mK} d^{\mJ}_{\mr\ms} =0 \; \Rightarrow \; h^{\mr}_{\mK} \big( T_{\mr} \big)^{\mJ}{}_{\mI}=0
\eea
that effectively imply $h^{\mr}_{\mI} T_{\mr} =0$ in any representation. Therefore, $T_{\mr}$ are not linearly independent and act as projectors, with $A_M^{\mr} T_{\mr}$ being a standard Yang-Mills field not subjected to any St\"uckelberg shifts, while the remaining part being purely St\"uckelberg. This prompts to split the indices $A_M^{\mr}\rightarrow \big(A_M^{\malpha}, A_M^{\ma} \big)$, $B_{MN}^{\mI} \rightarrow \big( B_{MN\ma{}^\prime}, B_{MN}^{\ma}\big)$, which was done in \cite{tenshier3}. The result of this analysis shows that one can choose $h^{\mr}_{\mI}$ in such a way that $h^{\ma}_{\mb}= \delta^{\ma}_{\mb}$ is its only nontrivial component; then $h^{\mr}_{\mI}g^{\ms\mI}=0$ implies that the only non-zero components of $g^{\mr\mI}$ are $(g_{\ma ^\prime}{}^{\malpha},g_{\ma ^\prime}{}^{\mb} ) = g_{\ma ^\prime}{}^{\mr}$. The algebra generators in both representations also split into doublets $T_{\mr} \rightarrow \big(T_{\malpha},T_{\ma} \big) $  and consistency relations imply that $T_{\ma}=0$ while $T_{\malpha}$ read
\be\label{genssplit}
T_{\malpha}^{(V)} = \left( \begin{array}{cc}
                           -f_{\malpha\mbeta}^{\mgamma} & -f_{\malpha\mbeta}^{\mb} +d_{\malpha\mbeta}^{\mb} \\
                            0                          &  \big( \widetilde{T_{\malpha}}  \big){}^{\mb}{}_{\ma}
                    \end{array} \right) ,
\;\; T_{\malpha}^{(T)} = \left(  \begin{array}{cc}
                           -g_{\mb{}^\prime}{}^{\mr} b^{\ma{}^\prime}_{\mr\ma} & 0\\
                            2 d_{\mb{}^\prime \malpha\ma} - g_{\mb^\prime}{}^{\mr} b_{\ma\mr\malpha}  &  \big( \widetilde{T_{\malpha}}  \big){}^{\mb}{}_{\ma}
                            \end{array}   \right)
\ee
in vector and tensor representations, respectively, and $\big(\widetilde{T_{\malpha}}\big){}^{\mb}{}_{\ma}$ form a Lie algebra
\be\label{lie2}
\big[  \widetilde{T_{\malpha}} ,  \widetilde{T_{\mbeta}}   \big] = f_{\malpha\mbeta}^{\mgamma} \widetilde{T_{\mgamma}} .
\ee

As $h^{\ma}_{\mb}= \delta^{\ma}_{\mb}$ and all other components of $h_{\mI}^{\mr}$ are zero, $A_{M}^{\ma}$ is purely St\"uckelberg field, which can be absorbed into $B_{MN}^{\ma}$ to make it invariant with respect to shifts. Tensor $B_{MN\ma{}^\prime}$ is also at least partially St\"uckelberg, as its transformation contains term $\delta B_{MN\ma{}^\prime} = - g_{\ma ^\prime}{}^{\mr}\Lambda_{MN\mr} + \ldots$. However, as at the same its rotations are generated by $-g_{\mb ^\prime}{}^{\mr} b^{\ma ^\prime}_{\mr \malpha}$ \p{genssplit}, precisely the  part of $B_{MN\ma{}^\prime}$ that experiences rotations is St\"uckelberg field, while the remaining part is Abelian. This leads to important conclusion about the tensor hierarchy:
it does not actually contain nontrivial generalizations of shift symmetries of Abelian antisymmetric tensors $\delta B_{MN} = \partial_M a_N - \partial_N a_M$. Instead only relevant transformations of $B_{MN}^{\ma}$ are only non-Abelian rotations and, if gauge group is not semi-simple, shifts involving Yang-Mills field strength:
\be\label{deltaB}
\delta B_{MN}^{\ma} = - \Lambda{}^{\malpha} \big( T_{\malpha} \big){}^{\ma}{}_{\mb} B_{MN}^{\mb} + \Lambda{}^{\malpha} \big(  -f_{\malpha\mbeta}^{\mb} +d_{\malpha\mbeta}^{\mb} \big) F_{MN}^{\mbeta}.
\ee
The field strength that is covariant with respect to these transformations is $F_{MNP}^{\ma}$ projection of $F_{MNP}^{\mI}$ \p{hierfieldstr}, and it does not involve 3-form $C_{MNK\mr}$. Therefore, non-trivial transformations of $C_{MNK\mr}$ do not play role in the hierarchy, and it can be assumed that it experiences gauge rotations only. Actually, Lagrangian considered in \cite{tenshier3} involves only the covariant part of $C_{MNP\mr}$ .

The 3-form $C_{MNK\mr}$ is still important for the construction \cite{tenshier3} as it acts as a Lagrange multiplier that induces self-duality equation on the field strength of $B_{MN}^{\ma}$. Using the split notation of the tensor hierarchy indices and spinor notation for spacetime ones, the Lagrangian \cite{tenshier3} reads
\be\label{hierlagr}
L_{TH} = -\frac{1}{4} \cD_{\alpha\beta} q_{\ma} \cD^{\alpha\beta} q^{\ma} -\im \chi_{\alpha\ma}^i\cD^{\alpha\beta}\chi_{\beta i}^{\ma} - \frac{1}{8}C_{(\alpha\beta){\ma}}\cD^{\alpha\gamma}B_{\gamma}{}^{\beta\ma} + \mbox{interactions}.
\ee
The conditions of existence of the Lagrangian \p{lagrcond} imply that $\big( T_{\malpha} \big){}^{\ma}{}_{\mb}$ and $\big( T_{\malpha} \big){}_{\mb^\prime}{}^{\ma^\prime} =  -g_{\mb{}^\prime}{}^{\mr} b^{\ma{}^\prime}_{\mr\ma} $ have to be contragredient representations, and one can contract respective indices $^{\ma}$, $_{\ma^\prime}$ via $\delta^{\ma^\prime}_{\ma}$. This means, however, that $q^{\ma}$ and $q_{\ma}= \delta^{\ma^\prime}_{\ma}q_{\ma^\prime}$ are fundamentally different objects, just as the fermions $\chi_{\alpha\ma}^i$, $\chi_{\beta i}^{\ma}$. This fits well with the superfield generalizations of the tensor hierarchy \cite{bandos,buchbinder} that employed two kinds of tensor multiplets. At the same time, however, this is the most obvious indication that the scalar kinetic term in the supersymmetric tensor hierarchy constructions \cite{tenshier2} has to possess indefinite metric. Also, $\chi_{\alpha\ma}^i$, $\chi_{\beta i}^{\ma}$ have to be superpartners of $C_{(\alpha\beta)\ma}$ and $B_\alpha{}^{\beta\ma}$ respectively and thus can not be linearly dependent.

The observations above suggest that choice of gauge transformations of the antisymmetric tensor is very limited and thus it is not necessary to consider the tensor hierarchy in full generality. Instead one may try to construct a system of the standard Yang-Mills field, antisymmetric tensor transforming similarly to \p{deltaB} and, possibly, other fields that are covariant with respect to the gauge group.
Moreover, according to the structure of generators \p{genssplit} nontrivial Chern-Simons-like couplings in the field strengths that involve $d^{\mI}_{\mr\ms}$ constants exist only if the gauge group is not semi-simple, i.e. not in the most interesting cases of $SU(N)$ and $SO(N)$. Therefore, we assume without much loss of generality that the non-Abelian tensor system within this approach contains only tensors without shift symmetries minimally coupled to the Yang-Mills fields and consider additional couplings only later in this work.

\subsection{Pasti-Sorokin-Tonin mechanism}
Observing the structure of the action \p{genssplit}, one can link the problem of indefiniteness of the metric in the scalar sector to the presence of the Lagrange multiplier that induces the self-dual equation of motion of the tensor field. As we already concluded, it is not actually needed from the gauge transformation point of view, and a possible solution to the problem of indefinite metric is to induce the tensor equation of motion in a different way.

The obvious alternative is the Pasti-Sorokin-Tonin mechanism \cite{PST}. The corresponding action involves, in addition to the antisymmetric tensor, one scalar field:
\bea\label{PSTact}
S_{PST} &=& \int d^6 x \Big(  \frac{1}{6}F_{MNP}F^{MNP} -\frac{1}{2 \partial^K z \partial_K z} \cF_{ABC}\cF^{ABD}\partial^C z \partial_D z    \Big), \nn \\
 \cF_{ABC} &=& F_{ABC} - \frac{1}{6}\epsilon_{ABCMNP}F^{MNP}, \;\; F_{ABC} = \partial_A B_{BC} - \partial_B B_{AC} + \partial_C B_{AB}.
\eea
This action possesses three gauge symmetries:
\bea\label{PSTtrans}
&& 1) \;\; \delta_f B_{MN}=\partial_M f_N - \partial_N f_M,\;\; \delta_f z=0, \nn \\
&& 2) \;\; \delta_\lambda z =\lambda, \;\;  \delta_\lambda B_{MN} = \frac{\lambda}{\partial^S z \,\partial_S z} \cF_{MNP}\partial^{P}z,\\
&& 3) \;\; \delta_a B_{MN} = \partial_{[M} z \; a_{N]}, \;\; \delta_a z=0. \nn
\eea
The first one is the usual gauge transformation of the antisymmetric tensor, the second implies that the scalar field is not an extra degree of freedom, and the last one is important to prove that the action \p{PSTact} produces self-dual equation of motion $\cF_{ABC}=0$.

The PST approach was combined with the tensor hierarchy in \cite{BSS}. Closer inspection shows, however, that it does not solve the problem. The constructed system employs similar hierarchy structure constants as \cite{tenshier2}, which already imply indefiniteness of the metric. The self-dual Lagrangian multiplier is still present and is actively used in proving invariance of the action with respect to one of PST gauge symmetries (shift of the auxiliary scalar by arbitrary function). Parameter of another transformation is constrained by $h^{\mr}_{\mI} a^{\mI}_M =0$, which, according to the splitting of the fields \cite{tenshier3}, implies that physical (i.e. not St\"uckelberg) part of the 2-form field is not actually transformed.

Using observations of the previous section, one can try to construct a much simpler system, considering just standard Yang-Mills field minimally coupled to a tensor $B_{MN}^{\mI}$:
\be\label{newB}
\delta A_{M}^{\mr} = \cD_{M}\Lambda^{\mr},\;\; \delta B_{MN}^{\mI} = - \Lambda^{\mr} \big(T_{\mr} \big){}^{\mI}{}_{\mJ} B_{MN}^{\mJ}.
\ee
Here, $\cD_{M} = \partial_M - A_M^{\mr} T_{\mr} \cdot$ and generators $T_{\mr}$ form a Lie algebra. Additionally, we assume that $\mI$ indices can be raised and lowered by the symmetric and positive-definite metric $\eta_{\mI\mJ}$, so that the generators with indices on one level are antisymmetric: $\big(T_{\mr} \big){}_{\mI\mJ}=-\big(T_{\mr} \big){}_{\mJ\mI}$. Then one can note that a straightforward non-Abelian generalization of the PST action
\bea\label{naivePST}
S_{NA} = \int d^6 x \Big(  \frac{1}{6}F_{MNP}^{\mI} F^{MNP}_{\mI} -\frac{1}{2 \partial^K z \partial_K z} \cF_{ABC}^{\mI}\cF^{ABD}_{\mI}\partial^C z \partial_D z    \Big),  \nn \\
F_{MNP}^{\mI}=\cD_{M}B_{NP}^{\mI}-\cD_{M}B_{NP}^{\mI}+\cD_{P}B_{MN}^{\mI}
\eea
does not possess even a symmetry with respect to linear transformations
\be\label{PSTtrans1}
\delta B_{MN}^{\mI} = \partial_M z \, a_N^{\mI} - \partial_N z \, a_M^{\mI},
\ee
which is obstructed by the term in the variation of \p{naivePST}
\be\label{obstr}
\frac{1}{18}\int d^6 x\,  \epsilon^{ABCMNP}\delta F_{ABC}^{\mI} \, F_{MNP\mI} =  \frac{1}{2}\int d^6 x\, \epsilon^{ABCDMN} F_{AB}^{\mr} \big( T_{\mr} \big)^{\mJ}{}_{\mI} B_{MN\mJ}\partial_C z \, a_D{}^{\mI}.
\ee
This problem can be solved by adding a term
\be\label{compterm}
 \frac{1}{2}\int d^6 x\, \epsilon^{ABCDMN} F_{AB}^{\mr} \big( T_{\mr} \big)^{\mJ}{}_{\mI} B_{MN\mJ}\partial_C z \, \frac{B_{DE}^{\mI} \partial^E z }{\partial_K z\, \partial^K z},\nn
\ee
which variation with respect to \p{PSTtrans1} exactly compensates \p{obstr}. This, however, complicates the equation of motion of $z$ and
makes generalization of the symmetry of the Abelian action
\be\label{PSTtrans2}
\delta z = \lambda, \;\; \delta B_{MN} = \lambda \frac{\cF_{MNP}\partial^P z}{\partial_K z\, \partial^K z}
\ee
impossible, so $z$ becomes a dynamical variable. More importantly, the modified equation of motion of $B_{MN}^{\mI}$
\bea\label{PSTBeom1}
0= \epsilon^{ABCMNP} \partial_P z \cD_C \Big(\frac{\cF_{MNK\mI}\partial^K z }{\partial_S z\, \partial^S z} \Big)   +\frac{1}{4}  \epsilon^{ABCDKL} F_{CD}^{\mr} \big( T_{\mr} \big)^{\mJ}{}_{\mI} B_{KL\mJ} - \\
- \frac{1}{2} \epsilon^{ABCDMN}F_{CD}^{\mr} \big( T_{\mr} \big)^{\mJ}{}_{\mI} \partial_M z \frac{B_{NK\mJ} \partial^K z }{\partial_S z\, \partial^S z} + \partial^{[A}z\, \epsilon^{B]KCDMN}F_{CD}{}^{\mr} \big( T_{\mr} \big)^{\mJ}{}_{\mI}\frac{B_{MN\mJ} \partial_K z }{\partial_S z\, \partial^S z}\nn
\eea
is not integrable as in the Abelian case, and it is impossible to conclude from it that $\cF_{MNP}^{\mI}=0$. If self-duality condition is enforced by hand, it would lead to further constraints on $B_{MN}^{\mI}$ (or $F_{MN}^{\mr}$).

\subsection{The bosonic action}
While straightforward generalization of the Pasti-Sorokin-Tonin action does not provide a workable solution, it still gives an idea how to circumvent the necessity to introduce two different physical tensor multiplets. Indeed, there exists a polynomial form of the PST action \cite{mkrtchyan}
\bea\label{PSTMact}
S_{PSTpoly} &=& \int d^6 x \Big(  \frac{1}{6}F_{MNP}F^{MNP} +\frac{1}{6}\big( \cF_{MNP} -3 \partial_{[M}z R_{NP]}  \big) \big( \cF^{MNP} -3 \partial^{[M}z R^{NP]}  \big)  \Big)= \nn \\
&=&  \int d^6 x \Big( \frac{1}{6}F_{MNP}F^{MNP} - \cF_{MNP}\partial^{M}z R^{NP} +\frac{3}{2}\partial_{[M}z R_{NP]}\partial^{[M}z R^{NP]} \Big).
\eea
which can be brought to the standard form by excluding auxiliary field $R_{MN}$ by its equation of motion. Superfield action of the tensor multiplet, which employs this mechanism to generate self-dual equation of the 2-form field, was constructed in \cite{susyPST}. Let us truncate action \p{PSTMact} by removing the last term, quadratic in $R_{MN}$ and $\partial_P z$:
\be\label{PSTredact}
S_{PSTtrunc} = \int d^6 x \Big( \frac{1}{6}F_{MNP}F^{MNP} - \cF^{MNP}\partial_{[P}z R_{MN]} \Big).
\ee
Note that the resulting action is still capable of producing self-dual equation of motion for $F_{MNP}$. Varying with respect to $R_{MN}$ one obtains equation $\cF^{MNP}\partial_{P}z =0$.\footnote{The same equation occurs as a part of proof that the PST action produces self-dual equation of motion.} Therefore, using anti-self-duality property of $\cF^{MNP}$,
\bea\label{cFeqsol}
\epsilon^{ABMNPQ}\cF_{MNP}\partial_Q z=0 \; \Rightarrow \; \cF_{MNP} = \partial_{[P}z K_{MN]} \; \Rightarrow \nn \\
 \big(\partial^P z\, \partial_P z \big)\, K_{MN} = \partial_M z\, \big( \partial^P z\, K_{NP}  \big) - \partial_N z\, \big( \partial^P z\, K_{MP}  \big).
\eea
If $\partial^P z\, \partial_P z \, \neq 0$, one can use the last relation to substitute $K_{MN}$  back and conclude that $\cF_{MNP}=0$. Moreover, using the same reasoning one can show that any self-dual tensor $C_{MNP}$ can be presented as
\be\label{Crep}
C_{MNP} = 3 \partial_{[M}z\, {\widetilde R}_{NP]} + \frac{1}{2}\epsilon_{MNPIJK}\partial^K z\,  {\widetilde R}^{IJ}.
\ee
Indeed, multiplying \p{Crep} by $\partial^P z$, one finds
\be\label{Crepsol}
{\widetilde R}_{MN} = - \frac{C_{MNP} \, \partial^P z}{ \partial^S z\partial_S z}+\mbox{irrelevant terms}\;\sim\;\partial z.
\ee
Then substitution ${\widetilde R}_{MN}$ \p{Crepsol} into \p{Crep} with use of relation $\epsilon_{[ABCMNP}\partial_{Q]}z=0$ results in an identity.

Formula \p{Crep} implies that the action \p{PSTredact} is equivalent to the standard one that induces self-dual equation of motion for $F_{MNP}$ with a Lagrange multiplier. It is different, however, from supersymmetric point of view, as it becomes possible to introduce a composite Lagrange multiplier without adding a new complete dynamical tensor multiplet. Therefore, appearance of the scalar component partner to the Lagrange multiplier, the key reason why indefinite metric appears in the tensor hierarchy actions, can be avoided.

Our approach to construct the non-Abelian tensor multiplet action can be, therefore, summarized in the following way. The starting point is the manifestly supersymmetric harmonic superspace action of the Abelian tensor multiplet constructed in \cite{susyPST}. We perform truncation of this action analogous to one that was done to the action \p{PSTMact} by removing one of the terms and add minimal interaction to the Yang-Mills multiplet. Then we add nonlinear conditions that allow us to remove excessive auxiliary components from the resulting action and show that the bosonic limit of the resulting action appears to be non-Abelian generalization of \p{PSTredact}. Analyzing the result, we discuss the possible constraints on the gauge group, demanding that the Lagrange multiplier has to be non-dynamical. Finally, we comment on the possibility of adding a non-minimal interaction.

\section{Improved supersymmetric action}
\subsection{Modifying supersymmetric PST action}
Truncation similar to one that was done to the action \p{PSTMact} can be performed in the manifestly supersymmetric way to the PST-type harmonic superfield action constructed in \cite{susyPST}
\bea\label{PSTsusy}
-8 S_{tensor}&=&  \int d^6 x d^4\theta^{-} du \Big[ D^{+}_\beta \Phi\big[ X\big]  \,D^{++}X^{+\beta} + \frac{1}{4}\Phi\big[ X\big] \,D^{++} D^{+}_\beta X^{+\beta} -  \\
&&-2 \Big(  D^{+}_\beta H\big[ Z,Y\big]  \,D^{++}X^{+\beta} + \frac{1}{4}H\big[ Z,Y\big] \,D^{++} D^{+}_\beta X^{+\beta}   \Big) + \nn \\
&&+ D^{++}Z \Big( D^{+}_\beta H\big[ Z,Y\big]  \,Y^{+\beta} + \frac{1}{4}H\big[ Z,Y\big] \, D^{+}_\beta Y^{+\beta}\Big) +\nn \\
&&+ M^{--} \big( D^{++}  \big)^3 Z + N^{+6} \big(  {D^{--}Z} + \im \frac{D^-_\alpha Z\, D^-_\beta Z\, \partial^{\alpha\beta}Z}{\partial_{\mu\nu}Z\partial^{\mu\nu}Z} \big) \Big].\nn
\eea
Note that in \cite{susyPST} two forms of the action are considered, one of which possesses symmetry $\delta Z = \Lambda$ at the superfield level and the other does not, relying on constraining $Z$ instead. We choose the form without explicit $\delta Z = \Lambda$ symmetry, as it becomes irrelevant after truncation.

The action \p{PSTsusy} is defined as an integral over analytic harmonic superspace \cite{HS,d6HS}, those properties are described in the Appendix A. The superfields $X^{+\alpha}$ and $Y^{+\alpha}$, satisfying
\be\label{tensormultdef}
D^+_\alpha X^{+\beta} = \frac{1}{4} \delta_\alpha^\beta D^+_\gamma X^{+\gamma}, \;\; D^+_\alpha Y^{+\beta} = \frac{1}{4} \delta_\alpha^\beta D^+_\gamma Y^{+\gamma}
\ee
describe the physical and auxiliary tensor multiplets, respectively, and contain $B$ and $R$ \p{PSTMact} tensors among their bosonic components
\be\label{tensormultcomp}
X^{+\alpha} = \theta^{+\beta} \big( \delta_\beta^\alpha\, q + B_\alpha{}^\beta   \big) + \ldots, \;\; Y^{+\alpha} = \theta^{+\beta} \big( \delta_\beta^\alpha\, c + R_\alpha{}^\beta   \big) + \ldots.
\ee

The superfield $Z$ is analytic, $D^+_\alpha Z =0$, and its first component is the PST scalar. Superfields $M^{--}$ and $N^{+6}$ are also  analytic and act as Lagrange multipliers to conditions that constrain components of $Z$ without putting it on shell. Quantities $\Phi\big[X\big]$ and $H\big[Z,Y\big]$ are defined as
\be\label{PhiHdef}
\Phi[X] = D^{--}D^+_\alpha X^{+\alpha} -2 D^-_\alpha X^{+\alpha}, \;\; H\big[Z,Y\big] = D^{--} Z\, D^+_\alpha Y^{+\alpha} -2 D^-_\alpha Z\, Y^{+\alpha}.
\ee
Analyticity of the Lagrangian in \p{PSTsusy} (and thus the possibility to integrate it over analytic superspace) is guaranteed by differential identities $\Phi\big[X\big]$ and $H\big[Z,Y\big]$ satisfy
\be\label{PhiHid}
D^+_\alpha D^+_\beta \Phi[X] =0,  \;\; D^+_\alpha D^+_\beta H\big[Z,Y\big] =0.
\ee

Keeping in mind that truncation of the bosonic action was performed by removing $R^2$ term, it is natural to remove
\be\label{removedterm}
 \int d^6 x d^4\theta^{-} du D^{++}Z \Big( D^{+}_\beta H\big[ Z,Y\big]  \,Y^{+\beta} + \frac{1}{4}H\big[ Z,Y\big] \, D^{+}_\beta Y^{+\beta}\Big)
\ee
from \p{PSTsusy}.

The minimal coupling to the non-Abelian gauge field can be achieved in standard way. We should consider $X$ and $Y$ as objects transforming with respect to some representation of the gauge group
\be\label{XYgaugetr}
\delta X^{+\alpha \mI} = -g \Lambda^{\mr} \big( T_{\mr} \big){}^{\mI}{}_{\mJ}X^{+\alpha \mJ}, \;\;  \delta Y^{+\alpha \mI} = -g \Lambda^{\mr} \big( T_{\mr} \big){}^{\mI}{}_{\mJ}Y^{+\alpha \mJ}.
\ee
Commutation relations of the generators $T_{\mr}$ and their action on various fields are taken as in \p{lie}, but the existence of a positive-definite non-degenerate metric $\eta_{\mI\mJ}$ is also assumed. $g$ is the coupling constant and $\Lambda^{\mr}$ is an analytic superfield parameter; the transformations \p{XYgaugetr} do not, therefore, contradict \p{tensormultdef}. Then one should promote the derivatives to covariant ones using appropriate superfields
\bea\label{Vs}
D^{++} \rightarrow \nabla^{++} = D^{++} - g V^{++\mr}  T_{\mr}\cdot, \;\;
D^{--} \rightarrow \nabla^{--} = D^{--} - g V^{--\mr} T_{\mr}\cdot, \nn \\
D^-_{\alpha} \rightarrow \nabla^{-}_\alpha = D^-_{\alpha}- g V^{-\mr}_{\alpha} T_{\mr}\cdot, \;\;
\partial_{\alpha\beta} \rightarrow \nabla_{\alpha\beta} = \partial_{\alpha\beta} - g V^{\mr}_{\alpha\beta} T_{\mr}\cdot,
\eea
with the derivative $D^+_\alpha = \partial/\partial \theta^{-\alpha}$ remaining covariant.
The gauge superfield $V^{++\mr}$ is analytic and transforms with respect to the gauge group as $\delta V^{++\mr} = \nabla^{++}\Lambda^{\mr}$. In the Wess-Zumino gauge it has the bosonic components
\be\label{Vpp}
V^{++\mr} \approx \theta^{+\mu}\theta^{+\nu}A_{\mu\nu}^{\mr} + \theta^{+4}C^{--\mr}, \;\; \partial^{--}A_{\mu\nu}^{\mr} = \partial^{--}C^{--\mr}=0,
\ee
with $A_{\mu\nu}^{\mr}$ being the standard Yang-Mills potential and $C^{--\mr} = C^{ij\mr}u^-_i u^{-}_j$ being the triplet of auxiliary fields.

Other superfields are not independent: $V^{--\mr}$ is related to $V^{++\mr}$ by equation
\be\label{Vmmdef}
\big[ \nabla^{++}, \nabla^{--}   \big] =D_0 \; \Rightarrow \; D^{++}V^{--\mr} - D^{--}V^{++\mr} - g f^{\mr}_{\ms\mt} V^{++\ms}V^{--\mt} = 0.
\ee
Superfields $V^{-\mr}_{\alpha}$, $V^{\mr}_{\alpha\beta}$ can be expressed in terms of $V^{--\mr}$
\be\label{YMSFVsol}
V^{-}_{\alpha} = -D^{+}_{\alpha}V^{--\mr}, \;\; V^{\mr}_{\alpha\beta} = \frac{\im}{2} D^+_\alpha D^+_\beta V^{--\mr}
\ee
so following (anti)commutation relations are satisfied
\be\label{YMSFcom}
\big\{ D^+_\alpha, \nabla^{-}_\beta\big\} = 2\im \nabla_{\alpha\beta}, \;\; \big[ \nabla^{++}, \nabla^{-}_{\alpha}   \big] =D^+_\alpha, \;\;\big[ \nabla^{--}, D^{+}_{\alpha}   \big] =\nabla^-_\alpha.
\ee
Other nontrivial commutation relations involve the super Yang-Mills field strength $V^{+\alpha\mr}$
\bea\label{YMSFcom2}
&&\big[  \nabla_{\alpha\beta}, D^{+}_\gamma \big] = \frac{\im g}{2} \epsilon_{\alpha\beta\gamma\mu}V^{+\mu\mr} T_{\mr}\cdot, \;\; \big[  \nabla_{\alpha\beta}, \nabla^{-}_\gamma \big] = \frac{\im g}{2} \epsilon_{\alpha\beta\gamma\mu}V^{-\mu\mr}T_{\mr}\cdot,  \\
&&V^{+\alpha\mr} =\big( D^{+3} \big)^{\alpha} V^{--\mr}, \;\;  D^+_\alpha V^{+\beta\mr} = \frac{1}{4}\delta_\alpha^\beta D^+_\gamma V^{+\gamma\mr}, \nn \\
&&\nabla^{++}V^{+\alpha\mr} =0, \;\; \nabla^{--}D^+_\alpha V^{+\alpha\mr} -2 \nabla^{-}_\alpha V^{+\alpha\mr} =0, \;\; V^{-\alpha\mr} = \nabla^{--}V^{+\alpha\mr}.\nn
\eea

Non-Abelian generalization of the truncated action is not as simple and straightforward as making all the derivatives involved covariant. The issue is that non-Abelian analog of $\Phi\big[X\big]$ \p{PhiHdef} satisfies the relation
\be\label{Phicovnot}
D^+_\alpha D^+_\beta \big( \nabla^{--}D^+_\gamma X^{+\gamma \mI} - 2\nabla^-_\gamma X^{+\gamma\mI}    \big) = -2 g \epsilon_{\alpha\beta\mu\nu} V^{+\mu\mr} \big( T_{\mr}  \big){}^{\mI}{}_{\mJ} X^{+\nu\mJ} \neq 0,
\ee
making straightforward generalization of the Lagrangian non-analytic. Our approach to deal with this problem is to use the quantity
\be\label{Phicov}
\Phi_{cov}\big[ X  \big] = \nabla^{--}D^+_\gamma X^{+\gamma \mI} - 2\nabla^-_\gamma X^{+\gamma\mI} + 2 g V^{+\gamma \mr} \big( T_{\mr}  \big){}^{\mI}{}_{\mJ}\, D^+_\gamma X^{--\mJ} - g D^+_\gamma V^{+\gamma \mr} \big( T_{\mr}  \big){}^{\mI}{}_{\mJ}\,  X^{--\mJ}.
\ee
Here $X^{--\mI}$ is the prepotential, $X^{+\alpha\mI} = \big(D^{+3}\big){}^{\alpha} X^{--\mI}$. $\Phi_{cov}\big[ X  \big]$ satisfies $D^+_\alpha D^+_\beta \Phi_{cov}\big[ X  \big]=0$ and reduces to $\Phi\big[X\big]$ \p{PhiHdef} when $g\rightarrow 0$, though this comes at the cost of introducing more components that are present in the $\theta-$expansion of $X^{--\mI}$.

With all these modifications made, the resulting naive covariantized truncated action \footnote{We rescaled $Y$ for simplicity.}
\bea\label{nonabnaive}
-8 S_{naive}&=&  \int d^6 x d^4\theta^{-} du \Big[ D^{+}_\beta\big( \Phi_{cov}\big[ X^{\mI} \big] +H\big[ Z,Y^{\mI}\big] \big) \,\nabla^{++}X^{+\beta}_{\mI} + \frac{1}{4}\big( \Phi_{cov}\big[X^{\mI}\big] +H\big[ Z,Y^{\mI}\big] \big) \,\nabla^{++} D^{+}_\beta X^{+\beta}_{\mI} \nn \\
&&+ M^{--} \big( D^{++}  \big)^3 Z + N^{+6} \left(  {D^{--}Z} + \im \frac{D^-_\alpha Z\, D^-_\beta Z\, \partial^{\alpha\beta}Z}{\partial_{\mu\nu}Z\partial^{\mu\nu}Z} \right) \Big]
\eea
is still not satisfactory. As analysis of equations of motion of \p{nonabnaive} shows, it contains many fields that are not to be expected in the tensor multiplet system. Some of these appear due to explicit presence of $X^{--\mI}$ in the action. Others result from harmonic expansions of usual physical fields, which are not constrained strongly enough. Of particular importance are the auxiliary field of the vector multiplet, $C^{--\mr}$, and the field $d^{--}_{\mu\nu}$ from expansion of $Z$
\be\label{Zexp}
Z = z + \theta^{+\mu}\theta^{+\nu} d^{--}_{\mu\nu} + \theta^{+4}e^{-4}
\ee
which induce couplings between the physical and auxiliary components and appear in the r.h.s. of the equation of motion of the physical scalar. We solve these problems by introducing nonlinear constraints on the superfields that do not put them on-shell but allow to remove unnecessary components, which come with new Lagrangian multipliers. The final superfield action reads
\bea\label{nonabfin}
-8 S_{NAT}&=&  \int d^6 x d^4\theta^{-} du \Big[ D^{+}_\beta\big( \Phi_{cov}\big[ X^{\mI} \big] +H\big[ Z,Y^{\mI}\big] \big) \,\nabla^{++}X^{+\beta}_{\mI} + \frac{1}{4}\big( \Phi_{cov}\big[X^{\mI}\big] +H\big[ Z,Y^{\mI}\big] \big) \,\nabla^{++} D^{+}_\beta X^{+\beta}_{\mI} \nn \\
&&+ M^{--} \big( D^{++}  \big)^3 Z + N^{+6} \left(  {D^{--}Z} + \im \frac{D^-_\alpha Z\, D^-_\beta Z\, \partial^{\alpha\beta}Z}{\partial_{\mu\nu}Z\partial^{\mu\nu}Z} \right) + L^{--}_{\mr\ms\mt} D^+_\alpha V^{+\alpha\mr}\,D^+_\beta V^{+\beta\ms}\,D^+_\gamma V^{+\gamma\mt} \Big]+ \nn \\
&&+ \int d^6 x d^8\theta du \, L^{-4}_{\mI\mJ}\nabla^{--}\nabla^{++}D^+_\alpha X^{+\alpha\mI}\,\nabla^{--}\nabla^{++}D^+_\beta X^{+\beta\mJ}.
\eea
\subsection{Bosonic limit}
Let us show that the action \p{nonabfin} possesses a sensible bosonic limit and no extra degrees of freedom from auxiliary superfields and harmonic expansions of dynamical fields appear. For this purpose, bosonic component expansions of superfields involved are needed. Most complicated is the unconstrained prepotential $X^{--\mI}$, which involves expansions in both $\theta^{-}$ and $\theta^{+}$ Grassmann coordinates:
\bea\label{Xmmexp}
X^{--\mI} &=& \cX^{--\mI} + \theta^{-\alpha}\cX{}^{-\mI}_{\alpha} + \theta^{-\alpha}\theta^{-\beta}\cX{}^{\mI}_{\alpha\beta} + \big( \theta^{-3} \big)_{\alpha}\cX{}^{+\alpha\mI} + \theta^{-4}\cX{}^{++\mI}, \\
\cX^{--\mI} &=& s^{--\mI} + \theta^{+\mu}\theta^{+\nu}s{}^{-4\mI}_{\mu\nu}+ \theta^{+4}s^{-6\mI}, \nn \\
\cX{}^{-\mI}_\alpha &=& \theta^{+\beta}t{}^{--\mI}_{\alpha\beta}+ \big( \theta^{+3} \big)_\beta t{}^{-4\beta\mI}_\alpha, \nn \\
\cX{}_{\alpha\beta}^{\mI} &=&  w_{\alpha\beta}^{\mI} + \theta^{+\mu}\theta^{+\nu}w_{\alpha\beta,\mu\nu}^{--\mI} + \theta^{+4}w_{\alpha\beta}^{-4\mI}, \nn \\
\cX{}^{+\alpha\mI}&=&  \theta^{+\beta}\big( \delta_\beta^\alpha q^{\mI} + B_\alpha{}^{\beta\mI} \big) + \big(\theta^{+3} \big)_{\beta}E^{--\beta\alpha\mI}, \;\; B_{\alpha}{}^{\alpha\mI}=0,\nn \\
\cX{}^{++\mI}&=&  f^{++\mI}+  \theta^{+\mu}\theta^{+\nu} a_{\mu\nu}^{\mI} + \theta^{+4}D^{--\mI}. \nn
\eea
Standard superfield potential $X{}^{+\alpha\mI} = \big(D^{+3}  \big){}^{\alpha}X{}^{--\mI}$ has the expansion $X{}^{+\alpha\mI} = \cX{}^{+\alpha\mI}+\theta^{-\alpha}\cX{}^{--\mI}$.

Bosonic expansions of $Y{}^{+\alpha\mI}$ and $Z$ are simpler:
\bea\label{YZexp}
Y{}^{+\alpha\mI} &=&  \theta^{+\beta}\big( \delta_\beta^\alpha c^{\mI} + R_\alpha{}^{\beta\mI} \big) + \big(\theta^{+3} \big)_{\beta}K^{--\beta\alpha\mI}+\theta^{-\alpha}\big( l^{++\mI}+\theta^{+\mu}\theta^{+\nu} b_{\mu\nu}^{\mI}+ \theta^{+4}J{}^{--\mI}   \big), \nn \\
Z &=&  z + \theta^{+\mu}\theta^{+\nu} d^{--}_{\mu\nu}+ \theta^{+4}e^{-4}.
\eea
Expansions of the Yang-Mills superfields $V^{++\mr}$ and $V^{--\mr}$ (which is a solution of equation \p{Vmmdef} for given $V^{++\mr}$) in the Wess-Zumino gauge read
\bea\label{Vmm}
V^{++\mr} &=& \theta^{+\alpha}\theta^{+\beta}A_{\alpha\beta}^{\mr} + \theta^{+4}C^{--\mr}, \;\; \partial^{--}A_{\alpha\beta}^{\mr} = \partial^{--}C^{--\mr}=0, \nn \\
V^{--\mr} &=& \theta^{-\alpha}\theta^{-\beta} U_{\alpha\beta}^{\mr} + \big( \theta^{-3} \big)_{\alpha} U^{+\alpha\mr}+ \theta^{-4}U^{++\mr}, \;\; \mbox{where}\nn \\
U_{\alpha\beta}^{\mr}&=& A^{\mr}_{\alpha\beta} - \frac{1}{12} \epsilon_{\alpha\beta\mu\nu} \theta^{+\mu}\theta^{+\nu} C^{--\mr},  \nn \\
U^{+\alpha\mr}&=&\theta^{+\beta} \big( 2\im F_\beta{}^{\alpha\mr} - \frac{1}{6}\delta_\beta^\alpha \partial^{++}C^{--\mr}  \big) +\frac{2\im}{3} \big( \theta^{+3} \big)_\beta \cD^{\beta\alpha}C^{--\mr}, \nn \\
U^{++\mr}&=& \frac{1}{6} \big( \partial^{++}  \big)^2 C^{--\mr} + \theta^{+\alpha}\theta^{+\beta} \Big( -\frac{\im}{6} \partial^{++}\cD_{\alpha\beta}C^{--\mr} -2\cD_{[\alpha\nu}F_{\beta]}{}^{\nu\mr}    \Big) +\nn \\
&&+ \theta^{+4} \Big( \frac{1}{3}\cD^{\mu\nu}\cD_{\mu\nu}C^{--\mr} + \frac{g}{6} C^{--\ms}\partial^{++}C^{--\mt}f_{\ms\mt}^{\mr} \Big) , \nn
\eea
where
\be\label{YMstr}
F_\beta{}^{\alpha\mr} = \partial_{\beta\mu}A^{\alpha\mu\mr} - \partial^{\alpha\mu}A_{\beta\mu}^{\mr} + \im g f_{\ms\mt}^{\mr} A_{\beta\mu}^{\ms}A^{\alpha\mu\mt}, \;\; \cD_{\mu\nu} = \partial_{\mu\nu} +\im g A_{\mu\nu}^{\mr} T_{\mr} \cdot
\ee
are the Yang-Mills field strength and the covariant derivative that acts on the components.

Expansions of the Lagrange multipliers  $L^{--}_{\mr\ms\mt}$, $L^{-4}_{\mI\mJ}$, $M^{--}$ and $N^{+6}$ appear to be not needed.

Actually, all the auxiliary fields can be removed by algebraic equations of motion obtained by varying \p{nonabfin} with respect to $Y^{+\alpha}$, $M^{--}$, $N^{+6}$, $L^{--}_{\mr\ms\mt}$ and $L^{-4}_{\mI\mJ}$.

Equations of motion of $M^{--}$ and $N^{+6}$ are the constraints on the superfield $Z$
\be\label{Zconstr}
\big( D^{++}  \big)^3 Z =0, \;\; {D^{--}Z} + \im \frac{D^-_\alpha Z\, D^-_\beta Z\, \partial^{\alpha\beta}Z}{\partial_{\mu\nu}Z\partial^{\mu\nu}Z} =0.
\ee
They were already solved in the bosonic limit in \cite{susyPST}. It was shown that \p{Zconstr} do not put $Z$ on-shell and restrict its component content to
\be\label{Zconstr2}
Z = z+\theta^{+\mu}\theta^{+\nu} d^{--}_{\mu\nu}, \;\; \partial^{--}z = \partial^{--} d^{--}_{\mu\nu}=0, \;\; \partial^{\mu\nu}z d^{--}_{\mu\nu} =\partial^{\mu\nu} d^{--}_{\mu\nu} =  d^{--\mu\nu} d^{--}_{\mu\nu}=0.
\ee
Then we may consider $L^{--}_{\mr\ms\mt}$ equation:
\be\label{Lmmeq1}
D^+_\alpha V^{+\alpha\mr}\,D^+_\beta V^{+\beta\ms}\,D^+_\gamma V^{+\gamma\mt}=0.
\ee
From \p{Vmm} it follows that
\bea\label{DV}
D^+_\alpha V^{+\alpha\mr} &=& \frac{2}{3} \big( \partial^{++}  \big)^2 C^{--\mr} + \theta^{+\alpha}\theta^{+\beta} \Big( -\frac{2\im}{3} \partial^{++}\cD_{\alpha\beta}C^{--\mr} -8\cD_{[\alpha\nu}F_{\beta]}{}^{\nu\mr}    \Big) +\nn \\
&&+ \theta^{+4} \Big( \frac{4}{3}\cD^{\mu\nu}\cD_{\mu\nu}C^{--\mr} + \frac{2g}{3} C^{--\ms}\partial^{++}C^{--\mt}f_{\ms\mt}^{\mr} \Big).
\eea
Therefore, the first component of equation \p{Lmmeq1} reads just
\be\label{Lmmeq2}
C^{++\mr}C^{++\ms}C^{++\mt}=0, \;\; C^{++\mr}\equiv \big(\partial^{++}\big)^2 C^{--\mr} = 2 C^{ij\mr}u^{+}_i u^{+}_j
\ee
and implies that in the bosonic limit $C^{ij\mr}=0$. With this taken into account, $D^+_\alpha V^{+\alpha\mr}$ reduces just to
\be\label{DV2}
D^+_\alpha V^{+\alpha\mr} =   -8\theta^{+\alpha}\theta^{+\beta} \cD_{[\alpha\nu}F_{\beta]}{}^{\nu\mr},
\ee
which cube is zero due to anticommuting properties of $\theta$'s. Therefore \p{Lmmeq1} only implies that $C^{ij\mr}=0$, without enforcing an equation of motion or otherwise constraining $F_\beta{}^{\alpha\mr}$. Note that the condition on $D^+_\alpha V^{+\alpha\mr} $ has to be at least cubic, as even quadratic equation would lead to constraint on $F_\beta{}^{\alpha\mr}$, while linear would also add to the action a dynamical Lagrange multiplier, leading to the kinetic terms with wrong signs for some fields.

Equation of motion of $L^{-4}_{\mI\mJ}$ is quadratic
\be\label{Lm4eq1}
\nabla^{--}\nabla^{++}D^+_\alpha X^{+\alpha\mI}\,\nabla^{--}\nabla^{++}D^+_\beta X^{+\beta\mJ}=0
\ee
and can be solved similarly to \p{Lmmeq1}. The bosonic $\theta$-expansion of $\nabla^{--}\nabla^{++}D^+_\alpha X^{+\alpha\mI}$ is rather long
\bea\label{nabla2DX}
&&\nabla^{--}\nabla^{++}D^+_\alpha X^{+\alpha\mI} = 4 \partial^{--}\partial^{++}f^{++\mI} + 4\im \theta^{-\alpha}\theta^{-\beta}\big( \partial_{\alpha\beta} + \im g U_{\alpha\beta}^{\mr}T_{\mr}\cdot  \big)\partial^{++}f^{++\mI} + \nn \\
&&+ 4g \big( \theta^{-3} \big)_{\alpha} U^{+\alpha\mr} \big( T_{\mr} \big){}^{\mI}{}_{\mJ} \partial^{++}f^{++\mJ} +4g \theta^{-4}U^{++\mr}\big( T_{\mr} \big){}^{\mI}{}_{\mJ}\partial^{++}f^{++\mJ}+\nn \\
&&+4 \theta^{+\mu}\theta^{+\nu}\partial^{--}\big( \im \cD_{\mu\nu}f^{++\mI} + \partial^{++}a_{\mu\nu}^{\mI}   \big) + 4\im \theta^{-\alpha}\theta^{-\beta}\theta^{+\mu}\theta^{+\nu}\big( \partial_{\alpha\beta} + \im g U_{\alpha\beta}^{\mr}T_{\mr}\cdot  \big)\big( \im \cD_{\mu\nu}f^{++\mI} + \partial^{++}a_{\mu\nu}^{\mI}   \big)+ \nn \\
&& + 8 \theta^{-\alpha}\theta^{+\beta} \big( \im \cD_{\alpha\beta}f^{++\mI} + \partial^{++}a_{\alpha\beta}^{\mI}   \big) +4g \big( \theta^{-3} \big)_{\alpha}\theta^{+\mu}\theta^{+\nu} U^{+\alpha\mr} \big( T_{\mr} \big){}^{\mI}{}_{\mJ}\big( \im \cD_{\mu\nu}f^{++\mJ} + \partial^{++}a_{\mu\nu}^{\mJ}   \big) +  \\
&& + 4g \theta^{-4}\theta^{+\mu}\theta^{+\nu}U^{++\mr}\big( T_{\mr} \big){}^{\mI}{}_{\mJ} \big( \im \cD_{\mu\nu}f^{++\mI} + \partial^{++}a_{\mu\nu}^{\mI}   \big)+ 4\theta^{+4}\partial^{--}\big( \partial^{++}D^{--\mI} -2\im \cD_{\mu\nu}a^{\mu\nu\mI}  \big) +\nn \\
&&+  4\im \theta^{-\alpha}\theta^{-\beta}\theta^{+4}\big( \partial_{\alpha\beta} + \im g U_{\alpha\beta}^{\mr}T_{\mr}\cdot  \big)\big( \partial^{++}D^{--\mI} -2\im \cD_{\mu\nu}a^{\mu\nu\mI}  \big)-4\theta^{-\alpha}\big( \theta^{+3} \big)_{\alpha}\big( \partial^{++}D^{--\mI} -2\im \cD_{\mu\nu}a^{\mu\nu\mI}  \big)+ \nn \\
&&+4g\big( \theta^{-3} \big)_{\alpha} \theta^{+4} U^{+\alpha\mr} \big( T_{\mr} \big){}^{\mI}{}_{\mJ}\big( \partial^{++}D^{--\mJ} -2\im \cD_{\mu\nu}a^{\mu\nu\mJ}  \big) +4g \theta^{-4}\theta^{+4}U^{++\mr}\big( T_{\mr} \big){}^{\mI}{}_{\mJ}\big( \partial^{++}D^{--\mJ} -2\im \cD_{\mu\nu}a^{\mu\nu\mJ}  \big).\nn
\eea
Quantities $U_{\alpha\beta}^{\mr}$, $ U^{+\alpha\mr}$, $U^{++\mr}$ are defined in \p{Vmm}, but their structure is not relevant here. Instead, one should note that the first component of the $\theta-$expansion of equation \p{Lm4eq1} reads
\be\label{Lm4eq2}
16 \partial^{--}\partial^{++}f^{++\mI}\partial^{--}\partial^{++}f^{++\mJ}=0,
\ee
which immediately implies $\partial^{--}\partial^{++}f^{++\mI}=0$, and, therefore, $\partial^{++}f^{++\mI}=0$ (see \cite{HS} for the rules of solving such equations). Removing $\partial^{++}f^{++\mI}$ from \p{nabla2DX}, one can obtain $\theta^{-2}\theta^{+2}$ term in the expansion
\be\label{Lm4eq3}
64 \theta^{-\mu}\theta^{+\nu}\theta^{-\rho}\theta^{+\sigma} \big( \im \cD_{\mu\nu}f^{++\mI} + \partial^{++}a_{\mu\nu}^{\mI}   \big)\big( \im \cD_{\rho\sigma}f^{++\mJ} + \partial^{++}a_{\rho\sigma}^{\mJ}   \big)=0.
\ee
One should take care factoring out Grassmann coordinates, as this induces some antisymmetry with respect to permutations of indices. Nevertheless, after setting $\sigma=\mu$ and $\rho=\nu$, one finds that
\be\label{Lm4eq4}
\big( \im \cD_{\mu\nu}f^{++\mI} + \partial^{++}a_{\mu\nu}^{\mI}   \big)\big( \im \cD_{\mu\nu}f^{++\mJ} + \partial^{++}a_{\mu\nu}^{\mJ}   \big)=0. \;\; \mbox{(no summation)}
\ee
Therefore, $\im \cD_{\mu\nu}f^{++\mI} + \partial^{++}a_{\mu\nu}^{\mI}=0$. Removing this combination from \p{nabla2DX} also, one sees that the remaining terms are proportional to at least $\theta^{+3}$ and make no contribution to the quadratic equation, leaving the quantity $\partial^{++}D^{--\mI} -2\im \cD_{\mu\nu}a^{\mu\nu\mI} $ unconstrained.

Finally, we should study equation obtained by varying the action \p{nonabfin} with respect to auxiliary tensor superfield $Y^{+\alpha}$ (or, more correctly, its prepotential $Y^{--\mI}$, $Y^{+\alpha\mI} = \big( D^{+3}\big)^{\alpha}Y^{--\mI}$). This equation and component equations that appear in its $\theta$-expansion, taking into account properties of $Z$ \p{Zconstr2}, read
\bea\label{Yeq}
&&H\big[ Z, \nabla^{++}X^{+\alpha\mI} \big] =0 \;\; \Rightarrow  \\
\theta^{+\alpha}\theta^{+\beta}:&& 4d_{\alpha\beta}^{--}\big( \partial^{++}q^{\mI}+ f^{++\mI}  \big) + 4 d^{--}_{[\alpha\gamma}\partial^{++}B_{\beta]}{}^{\gamma\mI}=0, \label{Yeq1} \\
\theta^{+4}:&& 4 d_{\alpha\beta}^{--}\big( -2\im\cD^{\beta\alpha}q^{\mI} -2\im \cD^{\gamma\alpha}B_{\gamma}{}^{\beta\mI} -2 a^{\beta\alpha\mI}+\partial^{++}E^{--\beta\alpha\mI}  \big)=0, \label{Yeq2} \\
\theta^{-\alpha}\theta^{+\beta}:&& 4d_{\alpha\beta}^{--}\partial^{++}f^{++\mI} -4\im \partial_{\alpha\beta}z\,\big( \partial^{++}q^{\mI}+ f^{++\mI}  \big) -4\im \partial_{\alpha\gamma}z\, \partial^{++}B_{\beta}{}^{\gamma\mI}=0,\label{Yeq3} \\
\theta^{-\alpha}\big(\theta^{+3}\big)_{\lambda}:&& -4\im \partial_{\alpha\beta}z \big(  -2\im\cD^{\beta\lambda}q^{\mI} -2\im \cD^{\gamma\lambda}B_{\gamma}{}^{\beta\mI} -2 a^{\beta\lambda\mI}+\partial^{++}E^{--\beta\lambda\mI} \big) - \label{Yeq4} \\
&&-8 d_{\alpha\beta}^{--}\big( \im \cD^{\beta\lambda}f^{++\mI} + \partial^{++}a^{\beta\lambda\mI} \big) + 8\im \partial_{\alpha\beta}d^{--\beta\lambda}\big( \partial^{++}q^{\mI}+ f^{++\mI}  \big) + 8\im \partial_{\alpha\beta}d^{--\gamma\lambda}\partial^{++}B_{\gamma}{}^{\beta\mI}=0.\nn
\eea

Consequences of all the mentioned equations should be studied together. As it is already known from \p{Lm4eq1} that $\partial^{++}f^{++\mI} =0$ and it is assumed $\partial^{\alpha\beta}z \partial_{\alpha\beta}z \neq 0$, one can obtain from \p{Yeq3}
\be\label{Yeq31}
\partial^{++}q^{\mI}+ f^{++\mI}=0, \;\; \partial^{++}B_{\beta}{}^{\gamma\mI}=0.
\ee
These algebraic equations relate $ f^{++\mI}$ to $q^{\mI}$ and show that the $B_{\alpha}{}^{\beta\mI}$ tensor has no component expansion. With them taken into account, equation \p{Yeq1} becomes trivial and \p{Yeq4} simplifies. To consider equation \p{Yeq4} properly, one has to take into account another consequence of \p{Lm4eq1}, particularly $\im \cD_{\mu\nu}f^{++\mI} + \partial^{++}a_{\mu\nu}^{\mI}=0$. Then $\partial_{\alpha\beta} z$ can be factored out, and \p{Yeq4} separates into three equations
\bea
&& -2\im\cD^{\beta\lambda}q^{\mI} -2\im \cD^{\gamma[\lambda}B_{\gamma}{}^{\beta]\mI} -2 a^{\beta\lambda\mI}+\partial^{++}E^{--[\beta\lambda]\mI}=0, \label{Yeq41} \\
&& \cD^{\gamma(\lambda}B_{\gamma}{}^{\beta)\mI}=0,\label{Yeq42} \\
&& \partial^{++}E^{--(\beta\lambda)} = 0.\label{Yeq43}
\eea
Equations \p{Yeq41} and \p{Yeq43} are purely algebraic. The first one can be used to obtain $a^{\mu\nu\mI}$, also making \p{Yeq2} satisfied.  The second just removes some of the fields. Equation \p{Yeq42} is the self-duality equation for the field strength of the $2-$form $B_\alpha{}^{\beta}$, the only dynamical equation in the whole system \p{Yeq}.

It is important to note that it is very difficult to obtain $a^{\mu\nu\mI}$  from equation \p{Yeq4} without using condition $\im \cD_{\mu\nu}f^{++\mI} + \partial^{++}a_{\mu\nu}^{\mI}=0$. As constraint $\partial^{++}f^{++\mI}=0$ can also be found as one consequences of $X^{--\mI}$ equation of motion, this is the only reason to add the last term to the action \p{nonabfin}.

If all algebraic equations of motion coming from \p{Lmmeq1}, \p{Lm4eq1}, \p{Yeq}
\bea\label{algeqs}
C^{++\mr}=0, \;\; \partial^{++}f^{++\mI}=0, \;\; \im \cD_{\mu\nu}f^{++\mI} + \partial^{++}a_{\mu\nu}^{\mI}=0, \;\; \partial^{++}B_{\alpha}{}^{\beta\mI}=0, \nn \\
\partial^{++}q^{\mI}+ f^{++\mI}=0, \;\; -2\im\cD^{\beta\lambda}q^{\mI} -2\im \cD^{\gamma[\lambda}B_{\gamma}{}^{\beta]\mI} -2 a^{\beta\lambda\mI}+\partial^{++}E^{--[\beta\lambda]\mI}=0
\eea
are taken into account, integration over anticommuting variables in \p{nonabfin} becomes straightforward. Terms with Lagrange multipliers $L^{-4}_{\mI\mJ}$, $L^{--}_{\mr\ms\mt}$, $M^{--}$ and $N^{+6}$ become zero identically, while $\nabla^{++}X^{+\alpha}_{\mI}$ reduces just to
\be\label{nablappX}
\nabla^{++}X^{+\alpha}_{\mI} \approx \big(\theta^{+3}\big)_{\lambda} \big( -2\im \cD^{\mu(\lambda}B_{\mu}{}^{\alpha)}{}_{\mI}  \big) + \theta^{-\alpha}\theta^{+4}\big( \partial^{++}D^{--}_{\mI} -2\im \cD_{\mu\nu}a^{\mu\nu}_{\mI}  \big).
\ee
As a result, most of the components present in the superfield prepotential $X^{--\mI}$ do not enter the action, with only the $\big(\theta^{-}\big)^{0}\big(\theta^{+}\big)^{0}$ and $\theta^{-\alpha}\theta^{+\beta}$ projections of ${\widetilde \Phi}{}_{cov}\big[X^{\mI}\big] + H\big[ Z,Y^{\mI}\big]$ making a contribution:
\bea\label{PhiHred}
{\widetilde \Phi}{}_{cov}\big[X^{\mI}\big] + H\big[ Z,Y^{\mI}\big] = 8 q^{\mI} + 4 \partial^{--}f^{++\mI} +\nn \\+ \theta^{-\alpha}\theta^{+\beta}\big[ 4 a_{\alpha\beta}^{\mI} - 4\im \cD_{\alpha\beta}q^{\mI} -4\im \cD_{\alpha\gamma}B_{\beta}{}^{\gamma\mI} +8\im g F_{\beta}{}^{\gamma\mr} \big( T_{\mr} \big){}^{\mI}{}_{\mJ}w_{\alpha\gamma}^{\mJ} -4\im \partial_{\alpha\beta}z c^{\mI} -4\im \partial_{\alpha\gamma}z\,R_{\beta}{}^{\gamma\mI}   \big]+ \nn \\+
\mbox{irrelevant terms.}
\eea
Substituting \p{nablappX} and \p{PhiHred} into \p{nonabfin}, one can perform integration over anticommuting variables. Assuming it is normalized as $\int d^4\theta^{-}  \theta^{+4} =1$, one obtains
\bea\label{nonabfin2}
-8S_{NAT} = \int d^6 x du \Big[ -8 \cD^{\gamma(\beta}B_{\gamma}{}^{\alpha)}{}_{\mI}\big( \cD_{\alpha\gamma}B_{\beta}{}^{\gamma\mI} -2 g F_{\beta}{}^{\gamma\mr} \big( T_{\mr} \big){}^{\mI}{}_{\mJ}w_{\alpha\gamma}^{\mJ}  +\partial_{\alpha\gamma}z\,R_{\beta}{}^{\gamma\mI}  \big) + \nn \\
  + 8 \big( q^{\mI} + \frac{1}{2} \partial^{--}f^{++\mI}\big) \big( \partial^{++}D^{--}{}_{\mI} -2\im \cD_{\mu\nu}a^{\mu\nu}{}_{\mI}  \big) \Big].
\eea
Integration over harmonic variables is also straightforward. One can note that as a consequence of \p{algeqs} combination $\tilde{q}^{\mI} \equiv q^{\mI} + \frac{1}{2} \partial^{--}f^{++\mI}$ does not depend on harmonics, just as $B_\alpha{}^{\beta\mI}$. As rules of harmonic integration \cite{HS} imply that $\int du 1 =1$ and integral of any traceless combination of harmonics is zero, $\int du$ effectively cuts off the harmonic dependence of multiples of $\tilde{q}^{\mI}$ and $B_\alpha{}^{\beta\mI}$. Finally, integrating by parts and substituting $a^{\mu\nu\mI}$ explicitly,
\bea\label{nonabfin3}
-8S_{NAT} = \int d^6 x \Big[ -8 \cD^{\gamma(\beta}B_{\gamma}{}^{\alpha)}{}_{\mI}\big( \cD_{\alpha\gamma}B_{\beta}{}^{\gamma\mI} -2 g F_{\beta}{}^{\gamma\mr} \big( T_{\mr} \big){}^{\mI}{}_{\mJ}w_{\alpha\gamma}^{\mJ}  +\partial_{\alpha\gamma}z\,R_{\beta}{}^{\gamma\mI}  \big)  + \nn \\ +16 \cD_{\mu\nu}\tilde{q}^{\mI}\cD^{\mu\nu}\tilde{q}_{\mI} + 8\im g \tilde{q}_{\mI}F_{\alpha}{}^{\beta\mr}\big( T_{\mr} \big){}^{\mI}{}_{\mJ}B_{\beta}{}^{\alpha\mJ} \Big],
\eea
where no field does depend on harmonics. Of a multitude of components of the prepotential $X^{--\mI}$, aside of the expected  $\tilde{q}^{\mI}$ and $B_\alpha{}^{\beta\mI}$, only $w_{\alpha\gamma}^{\mI}$ survives. Even this component is irrelevant, as it can be absorbed into Lagrange multiplier due to \p{Crep}. Actually, the self-dual part of $B$ field strength can also be absorbed, and one can write down the action analogous to the tensor hierarchy action \p{hierlagr}, but without a demand of a gauge group to be non-compact:
\bea\label{nonabfin4}
-8S_{NAT} = \int d^6 x \Big[ -4 \cD^{\gamma(\beta}B_{\gamma}{}^{\alpha)}{}_{\mI} C_{(\alpha\beta)}^{\mI} +16 \cD_{\mu\nu}\tilde{q}^{\mI}\cD^{\mu\nu}\tilde{q}_{\mI} + 8\im g \tilde{q}_{\mI}F_{\alpha}{}^{\beta\mr}\big( T_{\mr} \big){}^{\mI}{}_{\mJ}B_{\beta}{}^{\alpha\mJ} \Big], \nn  \\
C_{(\alpha\beta)}^{\mI} = 2 \cD_{(\alpha\gamma}B_{\beta)}{}^{\gamma\mI} -4 g F_{(\beta}{}^{\gamma\mr} \big( T_{\mr} \big){}^{\mI}{}_{\mJ}w_{\alpha)\gamma}^{\mJ} +2\partial_{(\alpha\gamma}z\,R_{\beta)}{}^{\gamma\mI}.
\eea

\subsection{Constraints on the group structure}
As this work was to large extent motivated by the necessity to find a non-Abelian tensor multiplet action without any fields having a kinetic term of a wrong sign, it is legitimate to ask a question whether the Lagrange multiplier introduces any complication, and in particular, is it dynamical at all. Indeed, one can vary the action \p{nonabfin4} with respect to tensor $B_{\gamma}{}^{\beta \mI}$ and
Yang-Mills $A_{\mu\nu}^{\mr}$ fields to obtain equations
\bea
\delta B_{\gamma}{}^{\beta \mI}: && 4\cD^{\gamma\rho}C_{(\rho\beta)\mI} = - 8\im g F_\beta{}^{\gamma\mr}\big( T_{\mr}  \big){}^{\mJ}{}_{\mI} \tilde{q}_{\mJ}, \label{Beq}  \\
\delta A_{\alpha\beta}^{\mr}: &&4\im g C_{[\alpha\gamma\mI} B_{\beta]}{}^{\gamma \mJ} \big( T_{\mr} \big){}^{\mI}{}_{\mJ} = -32\im g \big( T_{\mr} \big){}^{\mI}{}_{\mJ} \tilde{q}^{\mI}\cD_{\alpha\beta}\tilde{q}_{\mJ} +16\im g \big( T_{\mr} \big){}^{\mI}{}_{\mJ} \cD_{\gamma[\alpha}\big( \tilde{q}_{\mI}B_{\beta]}{}^{\gamma\mJ}  \big).\label{Aeq}
\eea
It can be noted that $C_{(\alpha\beta)\mI}$ enters equation \p{Aeq} without a derivative.  Acting by $\cD_{\alpha\gamma}$ on \p{Beq}, one obtains another algebraic equation \footnote{One can also apply $\cD^{\alpha\beta}$ to \p{Aeq} and then use \p{Beq}. The resulting expression vanishes if $\cD{}^{(\alpha\gamma}B_\gamma{}^{\beta)\mI}=0$ and $\tilde{q}{}^{\mI}$ equation of motion are taken into account.  }
\be\label{Beq2}
F_{[\alpha}{}^{\rho\mr} \big( T_{\mr} \big){}^{\mJ}{}_{\mI}C_{\beta]\rho\mJ} =4 \cD_{[\alpha\gamma}\big( F_{\beta]}{}^{\rho\mr} \big( T_{\mr} \big){}^{\mJ}{}_{\mI}  \tilde{q}_{\mJ}\big).
\ee
Equations \p{Aeq}, \p{Beq2} algebraically constrain $C_{(\alpha\beta)\mI}$, and, potentially, allow to express it in terms of other fields. If indices take values $\mr =1\,\ldots, r_m$ and $\mI =1,\ldots,I_m$, the self-dual tensor $C_{(\alpha\beta)\mI}$ has $4(4+1)/2 I_m = 10I_m$ independent components. Equations \p{Aeq} and \p{Beq2} impose at most $4(4-1)/2 r_m = 6r_m$ plus  $4(4-1)/2 I_m = 6I_m$ conditions on $C_{(\alpha\beta)\mI}$, although not all of them are necessary independent. Therefore, the tensor $C_{(\alpha\beta)\mI}$ can be completely defined by \p{Aeq}, \p{Beq2} only if number of equations is no less than a number of independent functions in $C_{(\alpha\beta)\mI}$, or
\be\label{groupcond}
6 r_m + 6 I_m \geq 10 I_m \;\; \Rightarrow \;\; 3r_m \geq 2I_m,
\ee
which puts a strong constraint on the acceptable representations of a gauged group with respect to which tensor multiplet transforms.

As one can note, even in simplest possible cases the system of \p{Aeq} and \p{Beq2} involves dozens of equations, and search for its analytical solution does not seem promising. However, one can study system \p{Aeq} and \p{Beq2} numerically for well known semi-simple  groups and their representations, assuming $B_{\alpha}{}^{\beta\mI}$ and $F_{\alpha}{}^{\beta\mr}$ are independent quantities. Relation \p{groupcond} provides a natural upper boundary for a dimension of acceptable representation for any given group, which simplifies the study. A note should be taken that the whole Lagrangian has to be real, and if $C$ is taken to be an inherently complex quantity, such as an $SU(N)$ spinor, it should always come with its Hermitean conjugate.

The numerical study shows that in all the cases when representation obeys dimensional restriction \p{groupcond} and does not contain a trivial component the homogeneous equations
\be\label{homeq}
F_{[\alpha}{}^{\rho\mr} \big( T_{\mr} \big){}^{\mJ}{}_{\mI}C_{\beta]\rho\mJ}=0, \;\; C_{[\alpha\gamma\mI} B_{\beta]}{}^{\gamma \mJ} \big( T_{\mr} \big){}^{\mI}{}_{\mJ} =0
\ee
have only identically zero solution. Also in all the cases with $3r_m > 2I_m$ non-homogenous system \p{Aeq}, \p{Beq2} has no solution for arbitrary right-hand side implying that it should be nontrivially constrained. In a very limited number of cases with $3r_m =2I_m$ system \p{Aeq}, \p{Beq2} has a solution that has no arbitrary functions and exists for any right-hand side of equations. These cases are
\begin{itemize}
\item The gauge group is $SU(3)$, $T_{\mr} \rightarrow T_{\malpha}{}^{\mbeta}$, $T_{\malpha}{}^{\malpha}=0$, $\malpha=1,2,3$, and $C$ is a pair of symmetric conjugated bispinors with upper and lower indices $C_{(\alpha\beta)}^{(\malpha\mbeta)}$, ${\overline C}_{(\alpha\beta)(\malpha\mbeta)}$.
\item The gauge group is $SO(4)$, $T_{\mr} \rightarrow T_{[\malpha\mbeta]}$, $\malpha=1,\ldots,4$. $C$ is a symmetric traceless tensor, $C_{(\alpha\beta)(\malpha\mbeta)}$, $C_{(\alpha\beta)(\malpha\mbeta)}\delta_{\malpha\mbeta}=0$.

\end{itemize}
For $Sp(2N)$ no representations with $3r_m = 2I_m$ were found. Possibly more solutions can be obtained by combining tensors within different representations or multiple copies within one representation.

The existence of solutions for only two groups is explained by the fact that number of components of multispinors and tensors rises strictly proportional to the dimension of the group only in the cases of traceless bispinor with one upper and one lower indices for $SU(N)$ and antisymmetric 2-tensor for $SO(N)$. In both cases it is simply equal to the dimension of the group, which is not acceptable. For all other spinors and tensors it rises slower or faster with $N$, and equality $3r_m =2I_m$ either is achieved only for certain $N$ or does not happen at all.

\subsection{Non-semi-simple groups}
The constructed action \p{nonabfin4} involves rather simple interaction terms, especially compared to the tensor hierarchy action \cite{tenshier3}, and it would be desirable to construct a more general system.  As was noted during the analysis of the tensor hierarchy, the possible transformations of the 2-form include gauge rotations and shifts by a term proportional to Yang-Mills field strength, if the gauge group is not semi-simple \p{deltaB}:
\be\label{deltaB2}
\delta B_{\alpha}^{\beta\mI} = - g \Lambda{}^{\mr} \big( T_{\mr} \big){}^{\mI}{}_{\mJ} B_{\alpha}^{\beta\mJ} - g \Lambda{}^{\mr} k_{\mr\ms}^{\mI} F_{\alpha}^{\beta\ms}.
\ee
Commutator of these transformations closes if \footnote{Note that the quantity $-f_{\malpha\mbeta}^{\ma} + d^{\ma}_{\malpha\mbeta}$ has to satisfy analogous relation to ensure that generators \p{genssplit} form a Lie algebra.}
\be\label{kprop}
- k^{\mI}_{\mt\ms}f^{\mt}_{\mr\mpp} - k^{\mI}_{\mpp\mt} f^{\mt}_{\mr\ms} - k^{\mJ}_{\mpp\ms} \big( T_{\mr}  \big){}^{\mI}{}_{\mJ}+ k_{\mr\mt}^{\mI}f^{\mt}_{\mpp\ms} + k^{\mJ}_{\mr\ms}\big( T_{\mpp}  \big){}^{\mI}{}_{\mJ}=0.
\ee
with $f^{\mt}_{\mr\ms}$ being structure constants of a Lie algebra.

Transformations \p{deltaB2} can be straightforwardly generalized to the superfield case, with $\Lambda{}^{\mr}$ now being an analytic superfield parameter
\bea\label{deltaX1}
\delta X{}^{--\mI} &=& - g \Lambda{}^{\mr} \big( T_{\mr} \big){}^{\mI}{}_{\mJ}X{}^{--\mJ} - g \Lambda{}^{\mr} k_{\mr\ms}^{\mI} V{}^{--\ms},\nn\\
\delta X{}^{+\alpha\mI} &=& - g \Lambda{}^{\mr} \big( T_{\mr} \big){}^{\mI}{}_{\mJ} X{}^{+\alpha\mJ} - g \Lambda{}^{\mr} k_{\mr\ms}^{\mI} V{}^{+\alpha\ms}.
\eea

Despite the fact that transformations of $B_{MN}^{\mI}$ in the tensor hierarchy are not homogeneous, a suitable covariant field strength can be found. In the Abelian superfield case, analogs of field strengths are the quantities $D^{++}X^{+\alpha}$ and $D^{--}D^+_\alpha X^{+\alpha} -2D^-_{\alpha}X^{+\alpha}$. It is not difficult to show that
\be\label{deltaDppX}
\delta \big( \nabla^{++}X^{+\alpha\mI} + g k^{\mI}_{\mr\ms} V^{++\mr}V^{+\alpha\ms}  \big) = -g \Lambda^{\mt} \big( T_{\mt} \big){}^{\mI}{}_{\mJ}\big( \nabla^{++}X^{+\alpha\mJ} + g k^{\mJ}_{\mr\ms} V^{++\mr}V^{+\alpha\ms}  \big)
\ee
with respect to transformations \p{deltaX1} if \p{kprop} and the Bianchi identity $\nabla^{++}V^{+\alpha\mr} =0$ are taken into account. Therefore, $\nabla^{++}X^{+\alpha\mI} + g k^{\mI}_{\mr\ms} V^{++\mr}V^{+\alpha\ms}$ transforms covariantly and is a correct generalization of $D^{++}X^{+\alpha}$ to the non-Abelian case. As one may expect, it satisfies
\be\label{DppXprop}
D^+_\beta \big( \nabla^{++}X^{+\alpha\mI} + g k^{\mI}_{\mr\ms} V^{++\mr}V^{+\alpha\ms}  \big) = \frac{1}{4} \delta^\alpha_\beta D^+_\gamma \big( \nabla^{++}X^{+\gamma\mI} + g k^{\mI}_{\mr\ms} V^{++\mr}V^{+\gamma\ms}  \big).
\ee

Generalization of $D^{--}D^+_\alpha X^{+\alpha} -2D^-_{\alpha}X^{+\alpha}$ is more subtle. As was shown in the case with minimal interaction, proper field strength that allows to construct an analytic superfield Lagrangian must involve terms with $V^{+\alpha\mr}$ and $X^{--\mI}$ in addition to the standard covariant ones \p{Phicov} so that $D^+_\alpha D^+_\beta \Phi_{cov}=0$. The first part of \p{Phicov} $\nabla^{--}D^+_\alpha X^{+\alpha\mI} -2\nabla^-_\alpha X^{+\alpha\mI}$ can be modified to be covariant with respect to transformations \p{deltaX1}
\bea\label{deltaDmmX}
\delta \big( \nabla^{--}D^+_\alpha X^{+\alpha\mI} -2\nabla^-_\alpha X^{+\alpha\mI} + g k_{\mr\ms}^{\mI} V^{--\mr}D^+_\alpha V^{+\alpha\ms} + 2g k_{\mr\ms}^{\mI} D^+_\alpha V^{--\mr}\, V^{+\alpha\ms} \big)=\nn \\
 = -g \Lambda^{\mt} \big( T_{\mt} \big){}^{\mI}{}_{\mJ} \big( \nabla^{--}D^+_\alpha X^{+\alpha\mJ} -2\nabla^-_\alpha X^{+\alpha\mJ} + g k_{\mr\ms}^{\mJ} V^{--\mr}D^+_\alpha V^{+\alpha\ms} + 2g k_{\mr\ms}^{\mJ} D^+_\alpha V^{--\mr}\, V^{+\alpha\ms} \big).
\eea
To prove \p{deltaDmmX}, one should use the relation \p{kprop} and the second Bianchi identity $\nabla^{--}D^+_\alpha V^{+\alpha\mr} -2\nabla^-_\alpha V^{+\alpha\mr}$. The requirement to cancel the derivatives of $\Lambda{}^{\mr}$ in \p{deltaDmmX} determines modification unambiguously. However, the second part of \p{Phicov} $2 g V^{+\gamma \mr} \big( T_{\mr}  \big){}^{\mI}{}_{\mJ}\, D^+_\gamma X^{--\mJ} - g D^+_\gamma V^{+\gamma \mr} \big( T_{\mr}  \big){}^{\mI}{}_{\mJ}\,  X^{--\mJ}$ can not be made covariant this way, and it appears to be impossible to find the second field strength that is both covariant and allows for construction of an action as an integral over analytic harmonic superspace.

\section{Conclusion}
In this article we constructed the harmonic superfield action with $N=(1,0)$, $d=6$ supersymmetry that involves non-Abelian extension of the tensor multiplet and possesses positive-definite metric in the scalar sector. Starting from the tensor hierarchy \cite{tenshier1,tenshier2}, we showed that possible non-trivial gauge symmetries of the tensor field include gauge rotations, and, if underlying group is not semi-simple, also shifts by the strength tensor of the Yang-Mills field while analogs of shift symmetries of the Abelian tensor are completely compensated by the St\"uckelberg fields. These transformations were realized in terms of harmonic superfields that define tensor multiplet off-shell.
The action we proposed was inspired by the Pasti-Sorokin-Tonin polynomial \cite{mkrtchyan}  and superfield \cite{susyPST} actions. While the standard PST mechanism does not function properly for the non-Abelian tensor fields, truncation of the corresponding action still induces the self-duality equation in the same way as the Lagrange multiplier. The difference from supersymmetric point of view is that in this case Lagrange multiplier is composite and can be introduced in such a way that does not involve extra dynamical scalar, allowing to keep the metric in the scalar sector definite. Analogous truncation of the action \cite{susyPST}, together with introduction of additional constraints to remove the auxiliary fields, forms the basis of our non-Abelian system. Minimal interaction with the Yang-Mills multiplet then can be introduced by promoting all the derivatives to the covariant ones using standard connection superfields, though modification of one of the tensor multiplet field strengths by terms explicitly involving prepotential is required to keep the superfield Lagrangian analytic. The component action was found in the bosonic limit. It involves the standard kinetic term for the scalar field and the Lagrangian multiplier for the self-duality equation of the tensor, as well as interaction of these fields with the Yang-Mills strength tensor. Conditions that make the Lagrangian multiplier non-dynamical were discussed, and two cases were found when this happens without algebraically constraining other involved fields. Study of possible modifications of the constructed Lagrangian showed that only semi-simple gauge groups are allowed.

Whole our construction was made possible by the use of harmonic superfield method that allowed us to deal with the multiplets and gauge symmetries completely off-shell and to avoid making any hypotheses about the on-shell transformation laws of physical fields, as is required in the component constructions. It is, obviously, desirable to compare our results to the standard supersymmetric tensor hierarchy ones \cite{tenshier2,tenshier3} in this context. However, current superfield construction involves a lot of nonlinear constraints on the auxiliary fields that have simple solutions only in the bosonic limit, and we prefer not to discuss the fermionic structures in this article.

\section{Appendix A}
The most important technique we use in this article is the harmonic superspace \cite{HS,d6HS}, which allows to formulate tensor and vector $N=(1,0)$, $d=6$ supermultiplets completely off-shell. This superspace involves, in addition to the usual space-time and Grassmann coordinates, the harmonics $u^{+i}$, $u^{-j}$, $i,j=1,2$, which satisfy
\be\label{harmdef}
u^{+i}u^{-j}\epsilon_{ij} =1, \;\; \epsilon_{ij} = -\epsilon_{ji}, \;\; \epsilon_{12}=1
\ee
and parameterize unit 2-sphere. Derivatives with respect to them are defined as
\be\label{harmder}
\partial^{++} = u^{+i}\frac{\partial}{\partial u^{-i}}, \;\; \partial^{--} = u^{-i}\frac{\partial}{\partial u^{+i}}, \;\; \partial_0 =  u^{+i}\frac{\partial}{\partial u^{+i}} -u^{-i}\frac{\partial}{\partial u^{-i}}
\ee
to preserve the relation \p{harmdef}.

In the harmonic superspace, the so called analytic basis can be chosen
\be\label{analyts}
x^{[\alpha\beta]}, \;\; \theta^{+\alpha}, \;\; \theta^{-\alpha}, \;\; u^{+i}, u^{-j},
\ee
in which one of the odd derivatives becomes simpler:
\bea\label{covdersanalyt}
D^{+}_\alpha = \frac{\partial}{\partial \theta^{-\alpha}}, \;\; D^{-}_\alpha = -\frac{\partial}{\partial \theta^{+\alpha}} -2\im \theta^{-\beta}\partial_{\alpha\beta}, \;\;  D_0 = \partial_0 + \theta^{+\gamma}\frac{\partial}{\partial \theta^{+\gamma}} -\theta^{-\gamma}\frac{\partial}{\partial \theta^{-\gamma}},  \nn \\
D^{++} = \partial^{++} + \im \theta^{+\alpha}\theta^{+\beta}\partial_{\alpha\beta} + \theta^{+\gamma}\frac{\partial}{\partial \theta^{-\gamma}}, \;\; D^{--}=\partial^{--} + \im \theta^{-\alpha}\theta^{-\beta}\partial_{\alpha\beta} + \theta^{-\gamma}\frac{\partial}{\partial \theta^{+\gamma}}.
\eea
The derivatives obey the (anti)commutation relations
\bea\label{hsderscom}
\big\{ D^{+}_\alpha, D^{-}_\beta  \big\}=2\im \partial_{\alpha\beta}, \;\; \big\{ D^{+}_\alpha, D^{+}_{\beta}\big\}=0, \;\; \big\{ D^{-}_\alpha, D^{-}_{\beta}\big\}=0, \nn \\
\big[ D^{++},D^{--}  \big]=D_0, \;\; \big[ D_0, D^{++} \big]=2D^{++}, \;\; \big[ D_0, D^{--} \big]=-2D^{--}, \nn \\
\big[ D^{++},D^{+}_\alpha   \big] =0, \;\; \big[ D^{--},D^{+}_\alpha   \big] =D^{-}_\alpha, \;\; \big[ D_0,D^{+}_\alpha   \big] = D^{+}_\alpha, \nn \\
\big[ D^{++},D^{-}_\alpha \big] =D^{+}_\alpha,  \;\; \big[ D^{--},D^{-}_\alpha   \big] =0, \;\; \big[ D_0,D^{-}_\alpha   \big] = -D^{-}_\alpha.
\eea
Simple structure of $D^+_\alpha$ allows to consider the analytic harmonic superspace $x^{[\alpha\beta]}, \;\; \theta^{+\alpha}, \;\; u^{+i},\;\; u^{-j}$ with twice less Grassmann coordinates.

Superfields defined on the harmonic superspace have to possess definite charge: $D_0 f^{q} = q f^{q}$. This reflects the fact that harmonics describe $S^2 = SU(2)/U(1)$, not whole $SU(2)$. Thus harmonic superfields are power series in $u^{+i}$, $u^{-j}$ with properly balanced charges. For example, for positive charge $q$
\be\label{harmsfexp}
f^{q}(x,\theta,u) = \sum^{\infty}_{n=0}f_{(i_1 \, \ldots \, i_{q+n}, j_1, \ldots, j_n)}(x,\theta) u^{+ i_1}\ldots u^{+i_{q+n}}  u^{-j_1} \ldots u^{-j_n}.
\ee
For $f^{q}$ with negative charge, roles of $u^{+i}$ and $u^{-i}$ are inverted. If net charge is zero, a harmonic-independent part may be present.
Integration over harmonic variables is defined by the rules
\be\label{harmintegr}
\int du \, 1 =1, \;\; \int du u^{+}_{(i_1} \ldots u^{+}_{i_n} u^{-}_{j_1} \ldots u^{-}_{j_m)} =0, \;\; m\; \mbox{or}\; n\neq 0.
\ee
In particular, $\int du \partial^{++}f^{--}(u)=\int du \partial^{--}f^{++}(u)=0$.

Integration over Grassmann coordinates in general and analytic superspaces is defined as
\be\label{grassint}
\int d^8\theta \,\theta^{-4}\theta^{+4} = 1, \;\; \int d^4\theta^{(-)}\theta^{+4} = 1.
\ee
Here $\theta^{-4}$, $\theta^{+4} $ are among the useful combinations of $\theta$s and derivatives, defined as \cite{buchbinder}
\bea\label{thetanot}
\big(\theta^{\pm 3}\big)_{\alpha} = \frac{1}{6}\epsilon_{\alpha\mu\nu\lambda}\theta^{\pm\mu} \theta^{\pm \nu} \theta^{\pm\lambda}, \;\;  \theta^{\pm4} = -\frac{1}{24}\epsilon_{\alpha\beta\mu\nu} \theta^{\pm\alpha}\theta^{\pm\beta}\theta^{\pm\mu}\theta^{\pm\nu}, \nn \\
\big(D^{\pm 3}\big)^{\alpha} = -\frac{1}{6}\epsilon^{\alpha\mu\nu\lambda}D^{\pm}_{\mu} D^{\pm}_{\nu} D^{\pm}_{\lambda}, \;\;  D^{\pm4} = -\frac{1}{24}\epsilon^{\alpha\beta\mu\nu}  D^{\pm}_{\alpha} D^{\pm}_{\beta} D^{\pm}_{\mu} D^{\pm}_{\nu}.
\eea

Throughout this article, we use spinor and vector notation interchangeably, as spinor one is more useful in the supersymmetric context and the vectors when the tensor hierarchy and the Pasti-Sorokin-Tonin action are discussed. The exact relation between these systems of notation is given by the gamma-matrices those properties can be found, for example, in \cite{buchbinder}. For our purposes it is sufficient to note that the vector $A_M$, $M=0,\ldots,5$ corresponds to the antisymmetric bispinor $A_{[\alpha\beta]}$, $\alpha=1,\ldots,4$, the antisymmetic 2-tensor $B_{MN}$ to the traceless bispinor $B_{\alpha}{}^{\beta}$, $B_{\alpha}{}^{\alpha}=0$, and self-dual and anti-self-dual 3-tensors can be written as $C_{(\alpha\beta)}$ and $C^{(\alpha\beta)}$, respectively. The spinor indices can be raised and lowered only as an antisymmetric pair
\be\label{spinrise}
A^{[\alpha\beta]} = \frac{1}{2}\epsilon^{\alpha\beta\mu\nu}A_{\mu\nu}.
\ee

\end{document}